\documentclass[notitlepage,twocolumn,prl]{revtex4-1}
\usepackage{amsmath}
\usepackage{mhchem} 
\usepackage{changes}
\usepackage{graphicx}
\usepackage[space]{grffile} 
\usepackage{gensymb}
\usepackage{bm}
\usepackage{amssymb,amsfonts,amsmath,amsbsy,bm,t1enc,lipsum,latexsym}

\begin{document}

\title{Probing the loss origins of ultra-smooth $\mathbf{Si_3N_4}$ integrated photonic waveguides}

\author{Martin H. P. Pfeiffer}
\author{Junqiu Liu}
\author{Arslan S. Raja}
\author{Tiago Morais}
\author{Bahareh Ghadiani}
\author{Tobias J. Kippenberg}
\email{tobias.kippenberg@epfl.ch}
\affiliation{{\'E}cole Polytechnique F{\'e}d{\'e}rale de Lausanne (EPFL), Laboratory of Photonics and Quantum Measurements (LPQM), Lausanne, CH-1015, Switzerland}

\date{\today}


\maketitle

\textbf{
	On-chip optical waveguides with low propagation losses and precisely engineered group velocity dispersion (GVD) are important to nonlinear photonic devices such as soliton microcombs, and likewise can be employed for on chip gyroscopes, delay lines or Brillouin lasers. Yet, despite intensive research efforts, nonlinear integrated photonic platforms still feature propagation losses orders of magnitude higher than in standard optical fiber.
	The tight confinement and high index contrast of integrated waveguides make them highly susceptible to fabrication induced surface roughness. Therefore, microresonators with ultra-high Q factors are, to date, only attainable in polished bulk crystalline, or chemically etched silica based devices, that pose however challenges for full photonic integration. Here, we demonstrate the fabrication of silicon nitride ($\mathrm{Si_3N_4}$) waveguides with unprecedentedly smooth sidewalls and tight confinement with record low propagation losses. This is achieved by combining the photonic Damascene process with a novel reflow process, which reduces etching roughness, while sufficiently preserving dimensional accuracy. 
	This leads to previously unattainable \emph{mean} microresonator Q factors larger than $5\times10^6$ for tightly confining waveguides with anomalous dispersion. Via systematic process step variation and two independent characterization techniques we differentiate the scattering and absorption loss contributions, and reveal metal impurity related absorption to be an important loss origin. Although such impurities are known to limit optical fibers, this is the first time they are identified, and play a tangible role, in absorption of integrated microresonators. Taken together, our work provides new insights in the origins of propagation losses in $\mathrm{Si_3N_4}$ waveguides and provides the technological basis for integrated nonlinear photonics in the ultra-high Q regime.
}

\section{Introduction}
Low propagation losses are a central requirement for planar on-chip optical waveguides in diverse application areas such as narrow linewidth lasers \cite{Belt2013}, integrated delay lines \cite{Lee2012a}, gyroscopes \cite{DellOlio2014} and quantum photonic circuits \cite{Politi2008}. Recently, the ability to achieve low loss and tight confinement waveguides, with precisely engineered dispersion properties, has become central to nonlinear integrated photonics, notably microresonator based optical frequency combs \cite{DelHaye2007, Kippenberg2011} operating in the dissipative Kerr soliton regime \cite{Herr2013a,Brasch2016} (soliton microcombs). Such soliton microcombs have enabled synthesis of broadband and coherent optical frequency combs with microwave line spacing in integrated devices, that have enabled counting of optical frequencies \cite{Jost2015, Brasch2017, Briles2017}, dual comb spectroscopy \cite{Suh2016}, terabit coherent communication \cite{Marin-Palomo2017}, ultrafast dual comb ranging \cite{Suh2017, Trocha2017}, low noise microwave generation \cite{Liang2015a}, astrophysical spectrometer calibration \cite{Obrzud2017,Suh2018}, and an all-photonic integrated frequency synthesizer \cite{Spencer2017}. 

Yet, while optical fibers with propagation losses below $0.5\,\mathrm{dB/km}$ form the backbone of today's global communication infrastructure, on-chip waveguides exhibit several orders of magnitude higher attenuation coefficients. The low losses of optical fibers were enabled by high-purity glasses developed in response to the seminal work of Kao that predicted low loss optical fibers when reducing impurities \cite{Kao1966}. So far, ultra-high Q microresonators could attain comparable values only when mitigating scattering losses via low confinement geometries or chemical polishing, in platforms such as silica wedges\cite{Lee2012}, or bulk crystalline resonators \cite{Ilchenko2004}. These platform are relying from a materials perspective on high purity glass, as used in optical fibers, or ultra-pure crystalline materials, originally developed for deep UV lithography \cite{Kippenberg2003, Savchenkov2004}. However, such platforms are not easily compatible with photonic integration: the low refractive index of silica waveguides requires an air-cladding, complicating photonic integration, while the fabrication of crystalline resonators is incompatible with common CMOS technology \cite{Yang2017}. As a consequence, materials that achieve similar levels of loss as silica, but with higher index for strong light confinement, could have significant benefit in the technological development of photonic integrated ultra-high Q microresonator technology. 

\begin{figure*}
	\includegraphics[width = \textwidth]{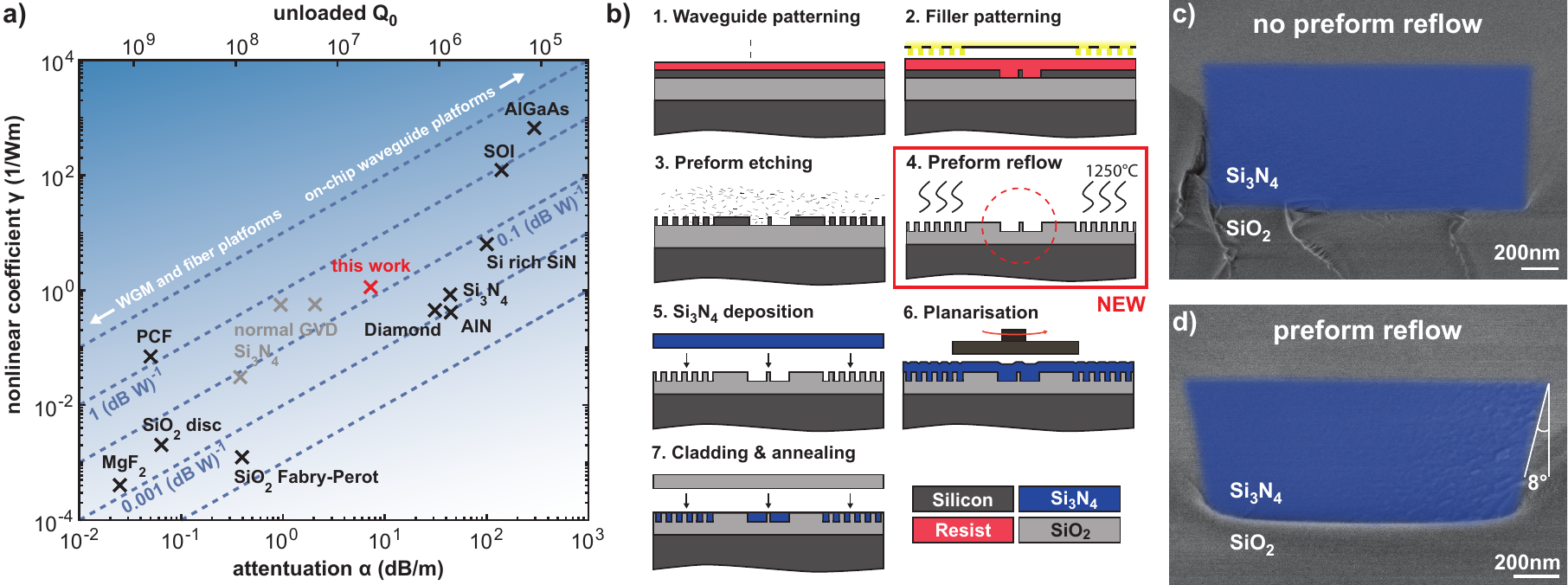}
	\caption{\textbf{Nonlinear waveguide platforms and photonic Damascene process with reflow step.} a) Overview of nonlinear coefficients $\gamma$ and attenuation $\alpha$ for different nonlinear waveguide platforms. Dashed lines indicate similar nonlinear performance based on a constant ratio of $\gamma/\alpha$. b) Schematic process flow of the photonic Damascene process highlighting the newly introduced preform reflow step, which consists in heating the substrate for sufficient time above its glass transition temperature. c,d) Final waveguide cross section from photonic Damascene process without and with reflow step. Rounding of the waveguide corners and an increased sidewall angle of $8^\circ$ are observed after reflow.}
	\label{fig_platformProcess}
\end{figure*}

Figure \ref{fig_platformProcess}a) compares the attenuation and nonlinear coefficient, $\alpha$ and $\gamma$, of the diverse low-loss waveguide and (ultra-)high-Q microresonator platforms. High index materials allow for tight confinement of light (small effective mode area $A_\mathrm{eff}$) and, following Miller's rule, a higher nonlinear refractive index $n_2$. Together, these features allow to attain effective nonlinear coefficients $\gamma = (2\pi n_2)/(\lambda A_\mathrm{eff})$ significantly higher than for low-loss crystalline or silica based platforms.  

In this context, silicon nitride ($\mathrm{Si_3N_4}$) waveguides \cite{Rahim2017} are an interesting compromise between nonlinearity, and propagation losses. First studied in the 1980'ies \cite{Henry1987} - it has emerged as a material platform for efficient nonlinear photonics, based on advances in processing that allowed overcoming the challenges of inherent film stress \cite{Moss2013}. The material system's wide transparency range and large bandgap of $\sim5\,\mathrm{eV}$ allows applications at visible wavelengths in the bio-medical context \cite{Hainberger2016}, telecom and mid-infrared wavelengths \cite{TaiLin2013}. Together, these properties allow the realization of tightly confining waveguides with high effective nonlinearity and precisely engineered, anomalous group velocity dispersion (GVD). The resulting low power threshold for parametric processes, has enabled the generation of broadband Kerr soliton frequency combs including dispersive waves \cite{Brasch2016} and coherent supercontinuum generation from high-repetition rate pulse trains \cite{Johnson2015}. 

Recent studies have shown that the intrinsic material absorption is compatible with propagation losses below of $1\,\mathrm{dB/m}$, and Q factors exceeding $2\times10^7$, at wavelengths around $1550\,\mathrm{nm}$ \cite{Ji2016, Xuan2016, Spencer2014}. However, the studies have not brought forward insights into where the residual losses are emanating from. Moreover, they were based on large waveguide geometries, chosen to limit scattering losses but incompatible with broadband dispersion engineering. To date, it  remains an outstanding challenge to fabricate $\mathrm{Si_3N_4}$ waveguides with reduced scattering losses, rivaling the nonlinear performance of silica wedge and polished crystalline microresonators. 

Here, building on recent advances of the photonic Damascene fabrication process \cite{Pfeiffer2016}, we demonstrate a methodology to create unprecedented smooth waveguides, that provide for the first time a platform in which (tightly confining) integrated photonics in the ultra-high Q regime can be made a reality. Importantly, our approach does not sacrifice mode confinement as in previous studies \cite{Bauters2011,Ji2016, Xuan2016}, and allows integrating resonators and bus waveguides within one material platform and layer. We perform an in-depth loss study for waveguides based on this novel process and are able to provide evidence that \emph{metal impurities} are an important origin of propagation losses. Although such impurities have been known to limit optical fibers, this is the first time they are identified, and play a tangible role, in the absorption of integrated photonic waveguides and microresonators. The photonic Damascene reflow process, as shown here, provides already record \emph{mean} Q-factors ($>5\times10^6$) for tightly confining ($2\times0.6\,\mathrm{\mu m}$) waveguides, which when combined with materials without impurities can attain values well in excess of $10^7$.

\section{Photonic Damascene process with preform reflow}
The photonic Damascene process solves several fabrication challenges of high-confinement $\mathrm{Si_3N_4}$ waveguides by inverting the common processing order, as schematically illustrated in Figure \ref{fig_platformProcess}b). Instead of etching the waveguide pattern into the highly stressed, micron thick $\mathrm{Si_3N_4}$ film, the latter is deposited onto a preform structured with recesses that form the waveguide pattern. A dense filler pattern surrounding the waveguide pattern relaxes the high tensile $\mathrm{Si_3N_4}$ film stress and prevents crack formation. Planar top surfaces, enabling heterogenous integration via bonding \cite{Chang2017} and with 2D materials, are prepared by removing the excess $\mathrm{Si_3N_4}$ using chemical mechanical polishing. The process enables the reliable fabrication of closely spaced high-confinement waveguides, which have successfully been applied in a growing number of nonlinear photonics experiments \cite{Trocha2017,Marin-Palomo2017,Herkommer2017, Pfeiffer2017a}.

As an addition to the above mentioned advantages, we incorporate a novel reflow step before the $\mathrm{Si_3N_4}$ deposition, aiming at smoothing the roughness of the waveguide preform. For this purpose, the wafer is heated slightly above the preform's glass transition temperature $T_\mathrm{G}$, in case of the wet thermally grown $\mathrm{SiO_2}$ preform about $1475\,\mathrm{K}$ \cite{Ojovan2008}. The surface-tension driven smoothing reduces especially the high spatial frequency components of the dry etch induced striations on the recess sidewalls. Here, we use a reflow process which consists of an 18h long anneal at $1523\,\mathrm{K}$ in oxygen atmosphere. As the temperature exceeds $T_\mathrm{G}$ only slightly, long annealing times are required. This is beneficial to control the reflow process and limit the changes in waveguide cross section. Nevertheless, as shown in Figure \ref{fig_platformProcess}c,d), an increased sidewall angle of $8^\circ$ and rounding of the waveguide corners is observed after the reflow.

\begin{figure*}
\includegraphics[width = \textwidth]{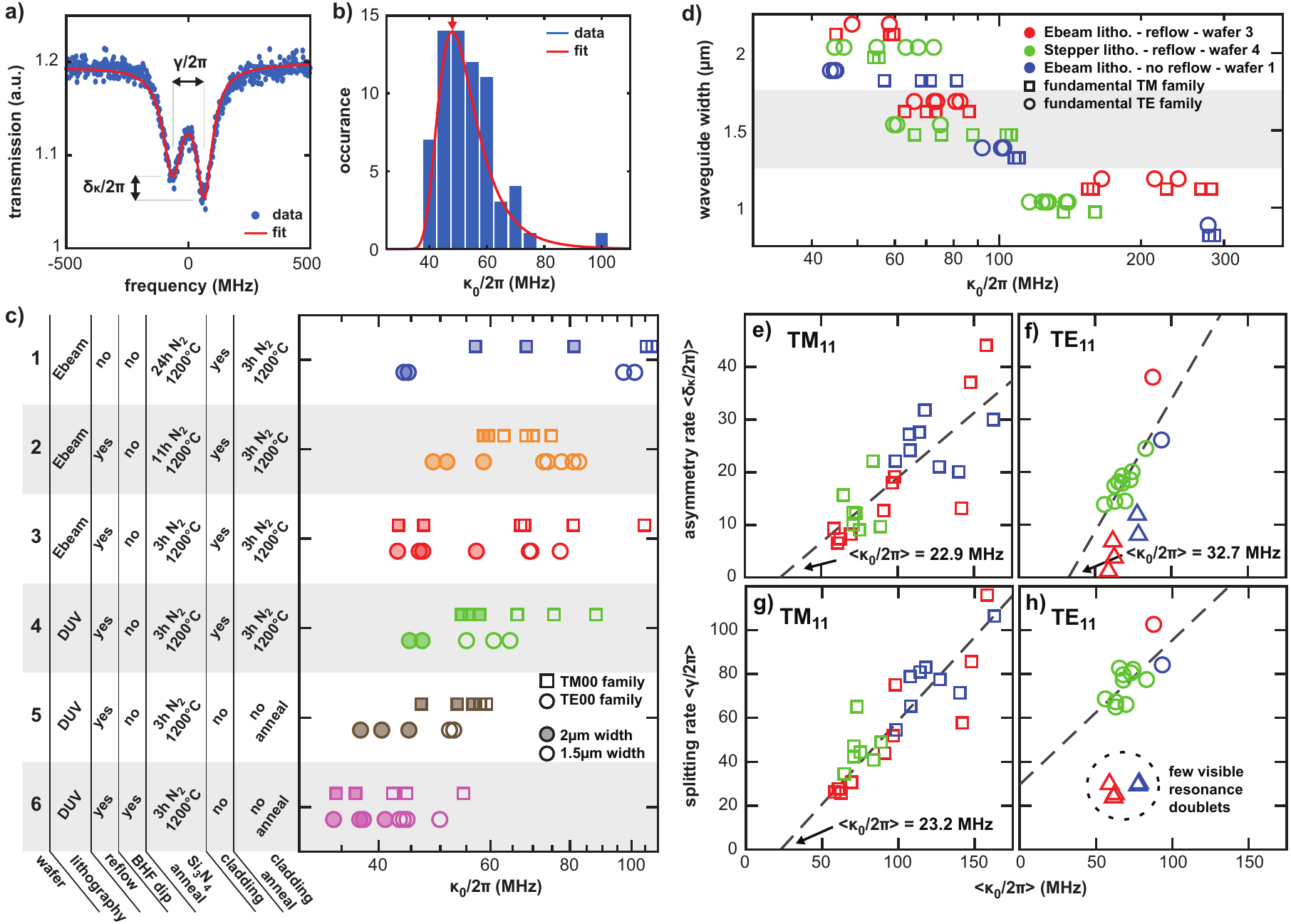}
\caption{\textbf{Systematic process comparison and resonance doublet analysis for 100-GHz FSR microresonators.} a) Resonance doublet with asymmetric linewidths fitted to extract the doublet asymmetry rate $\delta_\kappa/2\pi$ and the resonance splitting $\gamma/2\pi$, in addition to the intrinsic linewidth $\kappa_0/2\pi$. b) Histogram showing the occurrence of $\kappa_0/2\pi$ values within $5\,\mathrm{MHz}$ bins for the quasi-TM mode of a $2\,\mathrm{\mu m}$ wide microresonator from wafer 3. The distribution is fitted using a Burr distribution (red) to extract the most probable linewidth value (red arrow). c) Comparison of microresonator loss performance upon systematic variation of process parameters. d) Comparison of intrinsic loss rates for quasi-TE and -TM modes for microresonators with different widths, fabricated with and without reflow process. Improvements through the reflow process are visible for tightly confining waveguides with widths of $1.5\,\mathrm{\mu m}$ and smaller. e-h) Resonance doublet characteristics for $1.5\,\mathrm{\mu m}$ wide microresonators from different wafers. The mean coupling rate $\langle\gamma/2\pi\rangle$ and mean doublet asymmetry rate $\langle\delta_\kappa/2\pi\rangle$ are plotted for quasi-TM (e,g) and quasi-TE (f,h) polarized fundamental modes. Triangles indicate mean values based on less than 3 values. Linear correlations between the mean intrinsic loss rate and splitting or asymmetry rates are shown as dashed lines. The expected mean intrinsic loss rate for vanishing scattering is indicated.}
\label{fig_widthAnalysis}
\end{figure*} 

The ultra-low roughness of the $\mathrm{SiO_2}$ preform surface after the reflow step can hardly be perceived in scanning electron micrographs (see e.g Figure 2 in the SI). Atomic force microscopy (AFM) provides a method to measure such roughness to sub-nm levels and has been previously employed to asses the sidewall roughness of optical waveguides \cite{Poulton2006, Roberts2017}. To this end the AFM's tip is scanned along the waveguide sidewall, or in the present case along the waveguide recess' sidewall, as explained in detail in the SI. Denoting the local deviations from the waveguide dimensions by the random variable $f(z)$, the 1-D auto-correlation function $R(u_z) = \sum f(z)f(z+u_z)$ can be calculated. Common theoretical models estimate the scattering loss induced by the sidewall roughness based on the auto-correlation functions's RMS deviation $\sigma$ and correlation length $d$ \cite{Payne1994}. By fitting with a model composed of an exponential and a periodic part, we extract values of $\sigma = 0.18\,\mathrm{nm}$ and $d = 36\,\mathrm{nm}$ for the recess sidewall roughness after the reflow step. While the measured correlation lengths of the recess sidewall roughness is similar to previous works on Si and InP waveguides \cite{Borselli2004,Jang2003a, Poulton2006}, the estimated RMS deviation is significantly lower than previously reported values for $\mathrm{Si_3N_4}$ waveguides \cite{Roberts2017}.

\section{Systematic process comparison}
Next, we systematically compare the influence of the reflow step and other process variations on the propagation losses of high-confinement $\mathrm{Si_3N_4}$ waveguides. We infer the propagation losses by analyzing microresonator resonances whose linewidth $\kappa/2\pi$ is the sum of the coupling losses from the resonator to the bus waveguide $\kappa_\mathrm{ex}/2\pi$ and the internal losses $\kappa_0/2\pi$. The latter are related to the propagation coefficient $\alpha\,\mathrm{[m^{-1}]}$ as $\alpha = (n_\mathrm{eff}\kappa_0)/c$. The waveguide's core dimensions have strong influence on the propagation losses and, in contrast to previous works \cite{Xuan2016,Ji2016}, the here presented analysis focuses on dimensions allowing for broadband anomalous GVD at telecom wavelengths, an important property for nonlinear photonic applications. For our analysis we choose 100-GHz FSR microresonators with a radius of $\sim230\,\mathrm{\mu m}$, a core width of $1\,\mathrm{\mu m}$, $1.5\,\mathrm{\mu m}$ or $2\,\mathrm{\mu m}$ and a core height of $0.6-0.85\,\mathrm{\mu m}$.

Frequency calibrated transmission traces of microresonator devices are recorded for a wavelength range from $1500\,\mathrm{nm}$ to $1630\,\mathrm{nm}$ and processed using a setup and methods similar to the ones described in \cite{Pfeiffer2017a}. The resonances are automatically identified in the recorded transmission trace and the intrinsic loss rates $\kappa_0/2\pi$ are extracted through fitting of their lineshape. For this purpose either a Lorentzian lineshape model or a resonance doublet model are used. The latter can accurately fit resonance doublets with asymmetric lineshapes, as shown in Figure \ref{fig_widthAnalysis}a), and is discussed in detail in a later section. Based on their mutual FSR the resonances are manually grouped into mode families, which are identified through comparison with simulations. The resonator coupling regime is identified by comparing the trend of resonance linewidth to power extinction on resonance for resonators with the same geometry but varying resonator-bus waveguide distance on the same chip. The further analysis is based only on resonators in the under-coupled regime for which the intrinsic loss rate $\kappa_0$ dominates the overall cavity losses and coupling ideality related excess losses are low \cite{Pfeiffer2017a}. As the intrinsic linewidths can vary significantly across different resonances of the same microresonator, a single resonance measurement is not representative for a given device. In order to faithfully compare the performance of different fabrication processes, we therefore measure the statistical distribution of $\kappa_0/2\pi$ extracted from up to 150 resonances. As shown in Figure \ref{fig_widthAnalysis}b), the distribution can be well fitted using a Burr distribution \cite{Kordts2015} and its maximum is chosen as performance indicator, representing the \emph{most probable} propagation loss value for the microresonator device.

Figure \ref{fig_widthAnalysis}c) summarizes the loss performance observed for samples from 6 wafers with different process parameters. Per wafer the most probable loss value of the fundamental quasi-TE and quasi-TM mode families for several under-coupled 100-GHz FSR microresonators with $1.5\,\mathrm{\mu m}$ or $2\,\mathrm{\mu m}$ width are plotted. The adjacent table highlights the mutual processing differences among the wafers and further processing details of each wafer can be found in the SI. The process parameters were varied with the goal to observe trends originating from one of the usually suspected loss origins. Scattering losses due to sidewall roughness depend sensitively on the lithography and etching process used, and should be strongly reduced by the reflow step. Absorption losses are often associated with overtones of hydrogen impurities and high temperatue annealing steps of the $\mathrm{Si_3N_4}$ core as well as the $\mathrm{SiO_2}$ cladding films are known to reduce their residual hydrogen content.

The results in Figure \ref{fig_widthAnalysis}c) show no general trend for the loss relation between TE and TM polarized mode families, but the best values for both polarizations are achieved in the larger, $2\,\mathrm{\mu m}$ wide waveguides. The influence of annealing time as well as of the lithography process seems to be non-trivial, hinting to the fact that neither hydrogen related absorption nor sidewall roughness induced scattering losses have a clearly dominating role in the loss budget. Surprisingly, the comparison of wafers fabricated with and without reflow reveals little improvements. Similar observations are made when comparing different lithography processes and the best linewidths reached for most wafers is $\kappa_0/2\pi \approx 50\,\mathrm{MHz}$. This is different for cladding free devices, and reveals the negative influence of the low temperature oxide (LTO) cladding. The lowest \emph{mean} linewidths, smaller than $35\,\mathrm{MHz}$, are obtained when performing a short BHF dip directly after the reflow step. These values correspond to a resonator Q factors well above $5\times10^6$ and propagation coefficients of $\sim 5\,\mathrm{dB/m}$, record for waveguides with anomalous GVD.

Although unclear before, beneficial effects of the reflow step are revealed when comparing the loss performance of resonators with different widths. In Figure \ref{fig_widthAnalysis}d) the values for resonators with $1\,\mathrm{\mu m}$, $1.5\,\mathrm{\mu m}$ and $2\,\mathrm{\mu m}$ width fabricated either with or without reflow as well as different lithography methods are compared. As noted before, for $2\,\mathrm{\mu m}$ wide waveguides no performance difference is visible but clear improvements through the reflow process are visible for samples with $1\,\mathrm{\mu m}$ and $1.5\,\mathrm{\mu m}$ width. Moreover, when applying a preform reflow the loss performance appears to be independent of the lithography technique for samples with widths of $1.5\,\mathrm{\mu m}$ or larger.

A comparison of process step influence on loss performance is interesting, but unfortunately, the above presented data allows only indirect and ambiguous guesses of the propagation loss origins. In fact, knowledge of the relative strength of scattering and absorption losses is desirable to guide fabrication efforts. In the following we determine their relative fraction using two independent methods. First, we analyze the relation of linewidth and resonance doublet splitting to derive a limiting loss value in the absence of scattering losses \cite{Borselli2004}. Second, the measurement of the resonances' thermal bistability allows us to infer a spectrally resolved absorption loss rate \cite{Rokhsari2004}.

\section{Resonance doublet analysis}
Waveguide sidewall roughness couples the guided to radiation modes, causing scattering loss, and can moreover lead to a coherent build-up of the counter-propagating waveguide mode. A resonance doublet is observed if the coupling rate to the counter-propagating mode is similar or larger than the total resonance decay rate $\kappa$. The relation between scattering induced loss and reflection is non-trivial, but strongly correlated, and the analysis of resonance doublets is a common means to estimate scattering losses \cite{Little1996, Borselli2004, Kippenberg2009}. Most previous works used a simple coupled mode equation (CME) system with a real coupling coefficient $\gamma$ for the derivation of the splitted lineshape function \cite{Gorodetsky2000, Kippenberg2009}. However, the resulting expression does not provide accurate fitting for resonances doublets with unequal linewidths, as regularly observed in the context of high-confinement waveguide resonators \cite{Li2016, Li2013a, Borselli2004}.

In general the lineshape resulting from reflective scattering can vary strongly, and even resemble a Lorentzian with enlarged linewidth, depending on the relative position and amplitude of the participating scattering centers \cite{Zhu2010}. Therefore, Li et al. have proposed an extended CME model which includes also second-order coupling processes via radiation modes \cite{Li2013a}. The coherent (direct) and dissipative (indirect, via radiation bath) scattering processes are both loss-free but can interfere and thus yield a large variety of lineshapes including asymmetric resonance doublets. Here, we employ this extended model which practically entails a CME system with complex coupling coefficient $\kappa_\mathrm{c} = \kappa_\mathrm{c,R} + i\kappa_\mathrm{c,I}$:

\begin{equation}
\frac{\mathrm{d}a_m}{\mathrm{dt}} = -\left(i\Delta\omega+\frac{\kappa}{2}\right)a_m + i \frac{\kappa_c}{2} a_{-m} + \delta_\mathrm{m,CW}\sqrt{\kappa_\mathrm{ex}}s_\mathrm{in}
\label{eq_cmeModel}    
\end{equation}

with $a_m$ and $m, -m = \{\mathrm{CW,CCW}\}$ being the modal amplitudes of clockwise and counter-clockwise circulating modes, $\Delta\omega$ the cavity detuning and $s_\mathrm{in}$ the input laser driving the CW mode. As shown in Figure \ref{fig_widthAnalysis}a), the resulting lineshape function accurately fits the asymmetric resonance doublet and allows for the extraction of both the real and imaginary part of $\kappa_c$. The resulting total coupling strength, summing coherent (direct) and dissipative (indirect) coupling processes, can then be expressed as $\gamma = -\left(\kappa_\mathrm{c,R}/2\right)^2+\left(\kappa_\mathrm{c,I}/2\right)^2$. The doublet asymmetry rate, representing the competition between both coupling pathways and thus giving information about their relative strength, is expressed as $\delta_\kappa = (\kappa_\mathrm{c,R}\kappa_\mathrm{c,I})/2$.

We are interested in the effect of different lithography techniques and the reflow process on the scattering processes. To this end Figures \ref{fig_widthAnalysis}e-h) compare the characteristics of doublet lineshapes for $\mathrm{1.5\,\mu m}$ wide microresonators from the three wafers (1, 3 and 4) already analyzed for Figure \ref{fig_widthAnalysis}d). For each resonator, considering only their visibly splitted resonances, the mean value of the total coupling rate $\langle\gamma/2\pi\rangle$ and doublet asymmetry $\langle\delta_\kappa/2\pi\rangle$ is plotted as function of the mean intrinsic loss rate $\langle\kappa_0/2\pi\rangle$ for the fundamental quasi-TE and -TM modes. We note that the mean intrinsic loss rate is based only on the $\kappa_0/2\pi$ values extracted from fitting the resonance doublets, values obtained from resonances with Lorentzian lineshape are omitted. The resulting values are indicated as triangles if less than three splitted resonances were found for the resonator.

\begin{figure*}
\includegraphics[width = \textwidth]{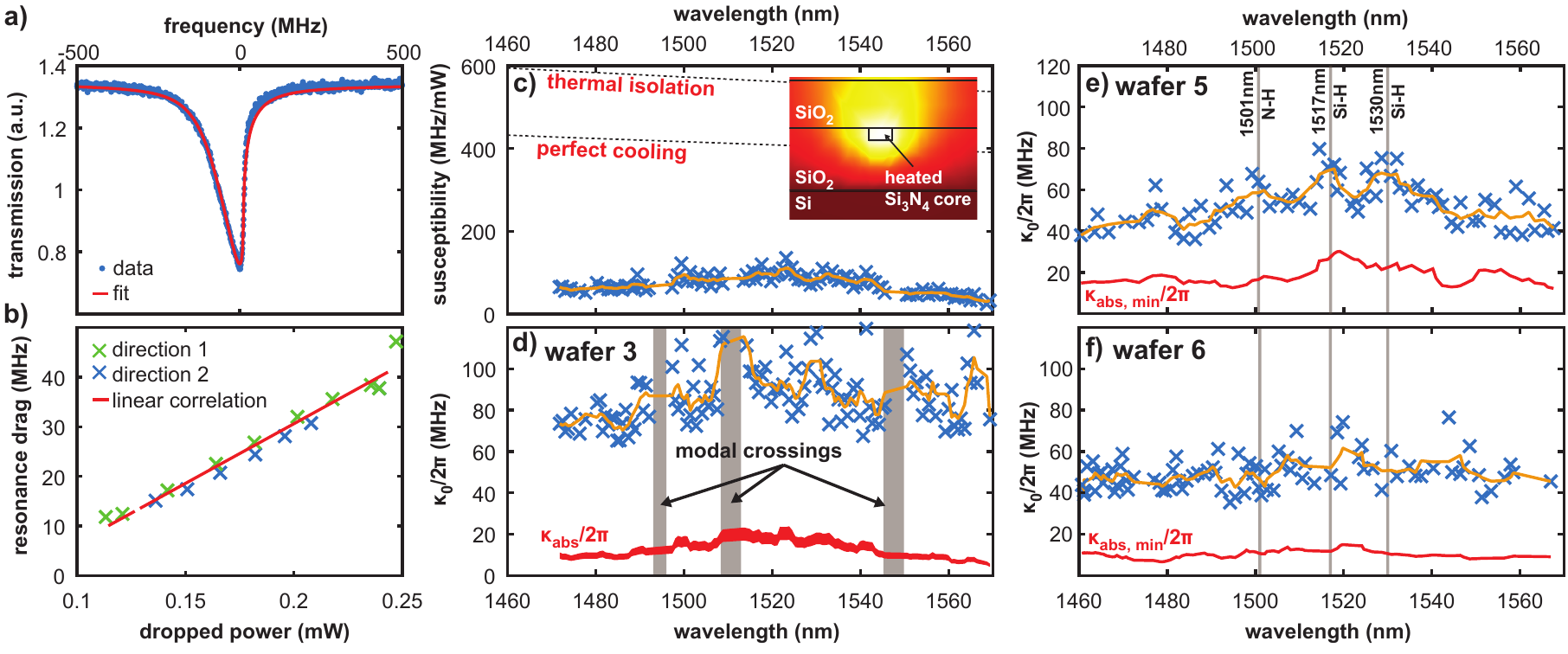}
\caption{\textbf{Determination of the absorption loss rate $\mathbf{\kappa_\mathrm{abs}}$ via thermal bistability spectroscopy of 100-GHz FSR microresonators with $\mathbf{1.5\,\mathrm{\mu m}}$ wide waveguides.} a) Measurement and fit of a skewed resonance lineshape due to the transient heating induced bistability. b) Linear correlation of extracted resonance drags for different dropped powers revealing the resonance's thermal susceptibility. c) Measured thermal susceptibilities of many resonances of a 100-GHz FSR microresonator from wafer 3. A moving average of the obtained values is superimposed in yellow. The estimated limits of the thermal susceptiblity in the case of complete absorption of the dropped power are shown as dashed lines. The inset shows an example of the simulated heat distribution for a uniformly heated waveguide core. d) Measured intrinsic linewidths $\kappa_0/2\pi$ and their moving average in yellow, corresponding to the thermal susceptibilities shown in c). The estimated absorption rate limits are shown in red. The positions of modal crossings leading to local deviations of the resonance properties are indicated in gray. e, f) Intrinsic loss rates measured for uncladded devices fabricated with and without 15s BHF dip, as well as the estimated lower limited of their absorption loss rates. The expected spectral position of hydrogen related overtone absorptions are shown in gray.}
\label{fig_dragSpectroscopy}
\end{figure*} 

Both the total coupling rate as well as the doublet asymmetry of the quasi-TM modes exhibit a clear correlation with the mean intrinsic loss rate. Independently, for both correlations an intrinsic loss rate of $\kappa_0/2\pi \approx 23\,\mathrm{MHz}$ for the case of vanishing scattering losses is extrapolated by fitting with a linear model. Moreover, the linewidth reduction through the preform reflow step can now be \emph{unambiguously} related to a reduced waveguide surface roughness. This becomes evident through the observed simultaneous reduction of resonance splitting and linewidth asymmetry rates with intrinsic loss rate. While the quasi-TM mode displays clear trends, the situation is more complex for the values obtained for the quasi-TE polarization. For samples fabricated using electron beam lithography with or without reflow step, only few resonances of each resonator show a visible splitting. This is different for resonators fabricated using stepper lithography for which most resonances of a given resonator exhibit visible splitting, with mean values about twice as large as found for the quasi-TM polarization. 

The latter is expected as the scattering for the quasi-TE polarization is dominated by the waveguide sidewall roughness which is higher than the waveguide's top and bottom surface roughness. In contrary, the little visibility of resonance doublets for the electron beam lithography samples is puzzling. The loss performance of $2\,\mathrm{\mu m}$ wide samples, as presented in Figure \ref{fig_widthAnalysis}c) and d), suggests that the losses other than scattering are similar for all three wafers. Moreover, for the non-reflown, $1.5\,\mathrm{\mu m}$ wide samples scattering rates higher than for reflown samples are expected and thus larger than the values of $\gamma/2\pi \approx 80\,\mathrm{MHz}$ obtained for wafer 4 (green) in Figure \ref{fig_widthAnalysis}h). In a simple scattering model the microresonators from the non-reflown wafer 1 (blue) should thus support many visibly split resonance doublets, as their expected intrinsic linewidth $\kappa_0/2\pi \approx 100\,\mathrm{MHz}$ (see Figure\ref{fig_widthAnalysis}c,d)) is on the same order. A similar reasoning can be made for wafer 3 (red), and overall we explain these observations by the non-trivial correlation between scattering losses and coherent reflection which depends on the statistical properties of the roughness, which are different for stepper or e-beam lithography. To extract a value for the intrinsic linewidth of the quasi-TE polarization in the absence of scattering losses, we base the correlation thus only on the values obtained for stepper lithography samples. However, a reasonable value of $\kappa_0/2\pi = 32\,\mathrm{MHz}$ is only obtained via the values of the resonance asymmetry $\delta_\kappa$.

In conclusion, we estimate a scattering loss rate of $\sim45\,\mathrm{MHz}$ for $1.5\,\mathrm{\mu m}$ wide waveguides when applying a preform reflow step, accounting for about two thirds of the total propagation losses. For larger waveguides lower values are expected and further experiments are needed to clarify the origin of the remaining losses.

\section{Thermal bistability spectroscopy}
Next, we perform systematic measurements of the resonance's thermal bistability to estimate the absorption loss rate \cite{Rokhsari2004, Borselli2005}. To this end the resonance frequency shift $\delta\omega$ as function of dropped power $P_d$ is measured, a quantity called thermal susceptibility $\chi_\mathrm{th}$ in the following. $\delta\omega$ and $P_d$ are related via the local temperature increase $\Delta T$ in the mode volume, which originates from the absorbed fraction $\zeta$ of the dropped power: $P_\mathrm{abs} = \zeta P_d$. The thermo-refractive effect and the thermal expansion of the mode volume, both described by the frequency dependent coefficient $\beta(\omega)$, cause the resonance frequency shift upon heating as $\delta\omega = \beta(\omega)\Delta T$. In practice, $\beta(\omega)$ can be easily measured by observing the resonance shift upon global heating of the sample, as described in detail in the SI. The structures thermal resistance $R_\mathrm{th}$ determines the local temperature increase upon heating with the absorbed power as $\Delta T = R_\mathrm{th} P_\mathrm{abs}$ and can be estimated via finite element simulations. Thus measuring $\chi_\mathrm{th}$ allows to calculate the absorption fraction $\zeta = \chi_\mathrm{th}(R_\mathrm{th}\beta(\omega))^{-1}$ which is alternatively expressed as absorption loss rate $\kappa_\mathrm{abs} = \zeta\kappa_0$.

The thermal susceptibility is measured by recording the skewed, triangular lineshapes of resonance upon red detuning a sufficiently intense laser across them, as shown in Figure \ref{fig_dragSpectroscopy}a). Such skewing originates from the thermal self-lock between driving laser and the resonance \cite{Carmon2004} and can be fitted with a steady-state model describing the bistability via a cubic equation to extract the associated resonance frequency shift $\delta\omega$. In order to precisely measure the dropped power on resonance, the frequency shift is measured for both on-chip coupling directions. As shown in Figure \ref{fig_dragSpectroscopy}b), a linear correlation of the measured resonance frequency shifts versus dropped powers allows then to determine the resonance's thermal susceptibility $\chi_\mathrm{th}$. Limiting values of the thermal resistance are simulated by considering a constant heating power in the core area and a fixed ambient temperature or perfect thermal isolation on the cladding top surface. This allows to calculate boundaries for $\chi_\mathrm{th}$, shown in Figure \ref{fig_dragSpectroscopy}, for the case of complete absorption of the power. Based on the measured and simulated values for the thermal susceptibility, as well as the measured intrinsic loss rates $\kappa_0/2\pi$, an absorption rate $\kappa_\mathrm{abs}/2\pi$ is then calculated. Details of the measurement setup, data processing and simulations can be found in the SI.

We measure $\chi_\mathrm{th}$ for quasi-TE polarized resonances of under-coupled 100-GHz FSR microresonators with $1.5\,\mathrm{\mu m}$ wide waveguides between $1460\,\mathrm{nm}$ and $1570\,\mathrm{nm}$. As shown in Figure \ref{fig_dragSpectroscopy}c) thermal susceptibilities around $80\,\mathrm{MHz/mW}$ are found for a fully cladded sample from wafer 3. With the estimated upper limits of $\chi_\mathrm{th}$ these values translate into a possible range for the absorption loss rate of up to $\kappa_\mathrm{abs}/2\pi \approx 20\,\mathrm{MHz} \pm 2\,\mathrm{MHz}$ between $1500\,\mathrm{nm}$ and $1540\,\mathrm{nm}$ and $\kappa_\mathrm{abs}/2\pi \approx 9\,\mathrm{MHz} \pm 2\,\mathrm{MHz}$ at the border of the measurement range. This result matches well with the residual, non-scattering losses found through the resonance doublet analysis for the quasi-TM modes. Figure \ref{fig_dragSpectroscopy}e,f) reveal similar absorption loss rates for uncladded samples from wafer 5 and 6, for which only a lower limit of $\kappa_\mathrm{abs}/2\pi$ (in the case of perfect top surface cooling) can be estimated.

The measurement range covers spectral regions ($1500\,\mathrm{nm}$ to $1540\,\mathrm{nm}$) where hydrogen impurity related absorption peaks are expected, as indicated in Figure \ref{fig_dragSpectroscopy}e,f) \cite{Germann2000,Bauters2011}. While in general the intrinsic loss rates $\kappa_0/2\pi$ (blue crosses) exhibit spectral variation, only in Figure \ref{fig_dragSpectroscopy}e) a slight increase of $\kappa_\mathrm{abs}/2\pi$ commensurate with the central Si-H related overtone is observed. We conclude that the absorption losses in our samples, especially in the best performing without top cladding, are dominated by a broadband absorbing species rather than the usually inculpated hydrogen impurities. Moreover, we find that the excess losses caused by the LTO cladding do not seem to be of absorptive nature, as the absorption loss rates of cladded and uncladded samples are very similar. In summary, absorption loss rates of $\sim20\,\mathrm{MHz}$ were found in the $1.5\,\mathrm{\mu m}$ wide waveguides, account for almost half the propagation losses. For wider waveguides a further reduction of the scattering loss contribution, and consequently a higher fraction of absorption losses in the total loss budget, is expected.

\section{Material analysis}

\begin{figure}
\includegraphics[width = \columnwidth]{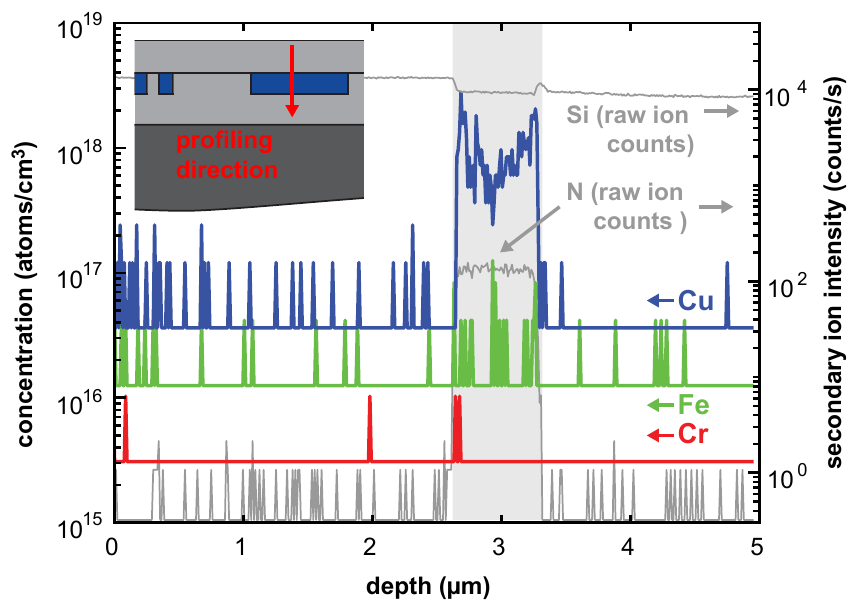}
\caption{\textbf{Concentration profile of common transition metal impurities in fully $\mathbf{SiO_2}$-cladded $\mathbf{Si_3N_4}$ sample.} Secondary ion mass spectroscopy (SIMS) allows to locally probed the metal concentration profile, as shown in the inset. The matrix raw ion counts of $\mathrm{Si}$ and $\mathrm{N}$ indicate the material layer composed of the top (LTO) and bottom (wet thermal oxide) cladding layers and the LPCVD $\mathrm{Si_3N_4}$ in between (gray background). The profiling is performed for copper, iron and chromium impurities, out of which the detected signal levels of iron and chromium are below the detection limit. A copper concentration of $\sim10^{18}\,\mathrm{atoms/cm^{3}}$ is measured within the $\mathrm{Si_3N_4}$ layer.}
\label{fig_metalImpurity}
\end{figure} 

Overtones of the optically active modes of Si-H and N-H bonds are the usual suspects for impurity related absorption losses in $\mathrm{Si_3N_4}$ waveguides \cite{Germann2000,Bauters2011}. The wavelength independent absorption loss rate that was found in the previous section brings this common knowledge into question for the here presented samples. Transition metal ions are an important class of impurities in the context of optical fibers, causing broadband absorption even at ppm-level concentrations \cite{Schultz1974}. Due to their efficient electronic trapping such impurities are also well-known in CMOS fabrication technology \cite{Graff2000}, e.g. in the context of solar cells \cite{Macdonald2004}, but to be best of our knowledge have never been considered in integrated photonics.

The precise measurement of ppm-level transition metal impurity concentrations is challenging and previously their concentration in LPCVD $\mathrm{Si_3N_4}$ thin films has been measured using vapor phase decomposition and X-ray fluorescence \cite{Vereecke2000a}. Here, we use glow discharge mass spectroscopy (GDMS) to analyze the concentration of common transition metals in samples of unprocessed $\mathrm{SiO_2}$ and $\mathrm{Si_3N_4}$ thin films. GDMS uses the rare gas ions created in a cathode discharge to sputter material of a surface. The sputter products are subsequently atomized in the glow discharge plasma, before entering a mass spectrometer. The technique offers impurity detection limits in the ppb range and does not suffer from matrix effects \cite{Harrison1986}.

Indeed, we measure concentrations between $0.1-2\,\mathrm{ppm wt}$ for transition metals, such as Cr, Fe and Cu, in all thin films that form the optical waveguide, even before processing. A detailed overview can be found in Table 2 in the SI. Typical processing steps such as dry etching and high temperature anneals can also  introduce impurities into the device, as well as cause their diffusive redistribution. To test the impurity levels in a final device and corroborate our findings, secondary ion mass spectroscopy (SIMS) is performed on fabricated samples to obtain quantified concentration profiles of the most prominent transition metal and hydrogen impurities. SIMS uses a localized ion beam to atomize material from the film stack which is subsequently analyzed in a mass spectrometer. A disadvantage of SIMS are so-called matrix effects which relate to the interaction between the ion beam and the matrix material, causing varying impurity extraction efficiency for different materials.

Figure \ref{fig_metalImpurity} shows the results obtained for a fully-cladded sample from wafer 3. Measurement details and further data for other samples are found in the SI. Neither chromium nor iron could be detected in concentrations above the respective detection limits but a copper concentration of  $\sim10^{18}\,\mathrm{atoms/cm^3}$ ($\approx 10\,\mathrm{ppm\,wt}$) is found in the $\mathrm{Si_3N_4}$ core area. Moreover, as shown in Figure 3 in the SI, hydrogen and chlorine impurities are found in concentrations of $\sim5\times10^{20}\,\mathrm{atoms/cm^3}$ ($\approx 5000\,\mathrm{ppm\,wt}$), respectively $\sim\times10^{19}\,\mathrm{atoms/cm^3}$ ($\approx 100\,\mathrm{ppm\,wt}$).

Based on these values, an exact derivation of the absorption losses induced by the impurities is difficult. Not only the values obtained by mass spectroscopy have significant error bars but also the absorbance of transition metal ions depends on their valence state which is generally unknown. For copper only the $\mathrm{Cu^{2+}}$ state is highly absorptive and thus literature values on absorption per ppm impurity concentration range from $0.1\,\mathrm{dB/km/ppm}$ to several hundreds $\mathrm{dB/km/ppm}$ \cite{Newns1973,Schultz1974,Ohishi1981}. We note that for an impurity concentration of $1-10\,\mathrm{ppm wt}$, as found for several transition metals in our samples, a value of $100\,\mathrm{dB/km/ppm}$ would equal to a significant broadband absorption loss rate of $\kappa_\mathrm{abs}/2\pi\approx10-100\,\mathrm{MHz}$ in the telecom C-band.

\section{Conclusion}
In summary we have presented a novel photonic Damascene reflow process enabling the fabrication of ultra-smooth waveguides. Using two independent characterization techniques we determined the scattering and absorption loss rates for tightly confining waveguide geometries, relevant for nonlinear photonics experiments. Our systematic study revealed a significant reduction of the scattering losses by the reflow process, resulting in dominant absorption losses in the best samples. A process study identified the cladding oxide as one main limiting factor for the current devices and \emph{mean} resonator Q-factors in excess of $5\times10^6$ in tightly confining waveguides are obtained. Moreover, for the first time, we were able to relate absorption losses in on-chip waveguides to the presence of transition metal ions. This is an important finding also relevant to the application of $\mathrm{Si_3N_4}$ waveguides in the visible range \cite{Rahim2017} where the absorption of most transition metal ions reached peak values \cite{Schultz1974}. Our results provide important insights into the losses origins of the widely used $\mathrm{Si_3N_4}$ waveguide platform. Based on this understanding future fabrication process improvements and new, advanced materials will enable on-chip microresonators for nonlinear applications with ultra-high Q-factors and propagation losses less than $1\,\mathrm{dB/m}$.

\vspace{3mm}

\emph{Data availability.} The code and data used to produce the plots within this paper are available at DOI:10.5281/zenodo.1169648. All other data used in this study are available from the corresponding authors upon reasonable request.

\subsection*{Acknowledgements}
$\mathrm{Si_3N_4}$ microresonator samples were fabricated in the EPFL Center of MicroNanotechnology
(CMi). This publication was supported by Contract HR0011-15-C-0055 from the Defense Advanced Research Projects Agency (DARPA), Defense Sciences Office (DSO). This work was supported by funding from the Swiss National Science Foundation under grant agreement No. 161573.

\bibliographystyle{unsrt}
\bibliography{lossPaperLiterature}

\end{document}